\documentstyle[prl,aps,multicol,psfig]{revtex}

\begin{document}
\title{Gap to Transition Temperature Ratio in Density Wave Ordering:
a Dynamical Mean Field Study}
\author{Stefan Blawid and Andrew Millis}
\address{Center for Materials Theory\\
Department of Physics \& Astronomy, Rutgers University\\
136 Frelinghuysen Road, Piscataway, NJ 08854}
\date{\today}
\maketitle

\begin{abstract}
We use the dynamical mean-field method to determine the origin of the large
ratio of the zero temperature gap to the transition temperature observed in
most charge density wave materials. The method is useful because it allows
an exact treatment of thermal fluctuations. We establish the relation of the
dynamical mean-field results to conventional diagrammatics and thereby
determine that in the physically relevant regime the origin of the large
ratio is a strong inelastic scattering.
\end{abstract}






\begin{multicols}{2}
Density wave ordering is a transition to a phase in which the
electronic charge or spin density has lower symmetry than the
underlying lattice. It occurs in a wide range of materials, including
quasi-one-dimensional organic conductors\cite{gruner94}, dicalogenides
such as ${\rm 2H-TaSe_{2}}$\cite {wilson75,mcwhan80}, `A-15' materials
like ${\rm V_{3}Si}$\cite{testardi75}, `blue bronzes' such as ${\rm
KMoO_{3}}$\cite{traviglini81}, and the `Verwey transition' material
${\rm Fe_{3}O_{4}}$\cite{coey98}. Much recent activity has related to
actual or possible 'stripe' density wave order in some members of the
high temperature superconductor family\cite{tranquada95} and to charge
and orbital order in some members of the `colossal' magnetoresistance
materials \cite{mori98}.

In most cases the density wave order evolves out of a metallic phase
and in this situation it is usually believed \cite{gruner94} 
to be driven by a
fermi surface 'nesting' instability. The resulting equations 
are similar to those of the `BCS' theory of
superconductivity and in particular lead to a ratio of $T=0$ density
wave gap $\Delta $ to density wave ordering temperature $T_{c}$ which
is close to the `BCS' value  $\Delta _{\rm BCS}/k_{\rm
B}T_{c}=1.76$. Almost all density wave materials, however, exhibit
much larger $\Delta /k_{{\rm B}}T_{c}$ ratios.  In
quasi-one-dimensional materials the large ratio may be understood
\cite{gruner94} as a consequence of critical fluctuations 
in low dimensionality, which decrease the
transition temperature more than
the $T=0$ gap $\Delta $. (Indeed, in a strictly one
dimensional material, $T_{c}=0$ while $\Delta > 0$). 
However, $\Delta /T_{c}$ values as
large as 10 are also observed in quasi two dimensional 
systems and in many fully three dimensional materials
\cite{wilson75,mcwhan80,testardi75,coey98}. 
These are not explicable either in the  BCS
approximation or in terms of the Migdal-Eliashberg-McMillan generalization
\cite{parks76} to so-called strong coupling superconductors such as Pb,
which exhibit $\Delta/T_{c}$ ratios only as large as 2.5. 

A generally accepted
explanation in the physically relevant regime
has not appeared. An extremely
large electron-phonon interaction could localize carriers as polarons
with a large activation gap; the ordering temperature $T_{c}$ would
then be controlled by a weak polaron-polaron repulsion 
and a large gap to $T_{c}$ ratio
would result. However, this would imply a 
non-metallic normal state with a resistivity which is large
and diverges rapidly as $T\rightarrow 0$, unlike the systems listed
above (except perhaps for $\rm Fe_{3}O_{4}$). 
McMillan made the intriguing suggestion (which has not
been followed up by subsequent workers as far as we know) that the low
$T_{c}$ (relative to $\Delta $) was a phonon entropy effect
\cite{mcmillan77}: thermal fluctuations of the
phonons would create areas where the local gap was small compared to
the average, and a self consistent process of exciting the electrons
into these areas would lead to the destruction of long ranged
order. 

In this paper we address the issue quantitatively using 
the dynamical mean-field method \cite{dmft} in the implementation
described in \cite{millis96}.
The technique  permits a complete solution
at all coupling strengths and temperatures,  
within a local (momentum independent self-energy)
approximation which is now generally accepted as reliable in $d=3$
spatial dimensions \cite{dmft}. We use it to show
that in the  weak to intermediate coupling regime
relevant to most of the materials listed above, 
the crucial physics is inelastic scattering of
electrons by the phonons involved in the lattice distortions. The
phonon entropy effects proposed by McMillan \cite{mcmillan77} are
present, but are
found be less important. This paper builds on 
the previous work of
Ciuchi and DiPasquale who extended the methods of 
\cite{millis96} to the density wave orderng case
and computed the phase diagram in the intermediate
to strong coupling regime \cite{ciuchi99}. We focus here on weak
to intermediate couplings and present new physical interpretations.

We study the Holstein model \cite{holstein59}, defined by  
\begin{equation}
H_{{\rm hol}}=-\sum_{ij}(t_{i-j}+\mu \delta _{ij})\,c_{i}^{\dagger
}c_{j}^{\phantom{\dagger}}+\frac{1}{2\lambda W
}\sum_{i}r_{i}^{\,2}+\sum_{i}r_{i} \left( c_{i}^{\dagger
}c_{i}^{\phantom{\dagger}}-n\right) \;.
\label{model}
\end{equation}

Here the operator $c_{i}^{\dagger }$ creates an electron on site $i$
and $r_{i}$ is the displacement of the ion on site $i$ measured from
the equilibrium displacement corresponding to to a uniform
distribution of electrons and rescaled so that the
electron-phonon coupling is unity. We have written the rescaled phonon 
stiffness in terms of an electron bandwidth parameter $W$ defined below
and a dimensionless coupling $\lambda$.
Because it contains all of the physics
of relevance to us and simplifies the computation substantially
we have adopted a classical limit (phonon kinetic term 
$\sum_{i}\frac{p_{i}^{2}}{2M}$ is
missing).  We comment below on the effects of the neglected quantum
fluctuations.   We also assume a bipartite lattice,
which for our purposes is a lattice possessing a dispersion
$\epsilon_{\vec{k}}$ (fourier transform of $t_{i-j}$) and a
wavevector $\vec{Q}$ such that $\epsilon _{\vec{k}-\vec{Q}}=-\epsilon
_{\vec{k}}$ for all $\vec{k}$ in the Brillouin zone. An equivalent
definition is that the lattice may be divided into two sublattices $A$
and $B$, such that $t_{i-j}$ only connects $A$ sites to $B$ sites.
We further specialize to
$n=1/2$; this implies $\mu =0$.

At  $n=1/2$ the ground state of  (\ref{model}) may
be shown to be charge ordered for all $\lambda $ by minimizing 
$\langle H_{{\rm hol}}$ $\rangle $ over
$r_{i}$. To compute $\langle H_{{\rm hol}}$ $\rangle $ we introduce
operators $a^{\dagger }$ and $b^{\dagger }$ creating electrons on the
$A$ and $B$ sublattices respectively and make the ansatz that
$r_{i}=\Delta $ on the A sublattice and $r_{i}=-\Delta $ on the B
sublattice. In momentum space the Hamiltonian may be written as the
$2\times 2$ matrix $H_{{\rm hol}}\left( \Delta \right) =
\frac{1}{2}\,N\,\frac{\Delta ^{2}}{\lambda
}+\frac{1}{2}\,\sum_{\vec{k}}\left( a_{\vec{k}}^{\dagger
}\;b_{\vec{k}}^{\dagger }\right) \,\left[ \Delta \,{\bf \tau _{z}}+\epsilon
_{\vec{k}}\,{\bf \tau _{x}}\right] {a_{\vec{k}} \choose b_{\vec{k}} }$where
$\tau _{z}$ and $\tau _{x}$ are the usual Pauli matrices
and all k-sums are over the full zone.  The
resulting energy dispersion is $E_{\vec {k}} = \pm \sqrt{{\epsilon
_{\vec{k} }^{2}+\Delta ^{2}}}$ and the energy of ${\rm H_{hol}\left(
\Delta \right) }$ may easily be found, and minimized over $\Delta
$. The equation yielding the minimum value is
\begin{equation}
1=\lambda W \,\frac{1}{N}\,\sum_{\vec{k}}\frac{1}{\sqrt{\epsilon _{\vec{k}
}^{2}+\Delta ^{2}}}\;.  \label{gap}
\end{equation}
If $\mu = 0$ the integral, (\ref{gap}), is logarithmically divergent 
as $\epsilon_k \rightarrow 0$
and thus a 
$\Delta >0$ solution exists. Later in this paper we use the
semicircular density of states $\rho _{0}(\epsilon
)=\frac{1}{N}\,\sum_{\vec{k}}\delta (\epsilon -\epsilon _{
\vec{k}})=\frac{2}{\pi \,W^{2}}\sqrt{W^{2}-\epsilon ^{2}}$ for which the equation
can be solved analytically, yielding  
$\Delta =\frac{4\,W}{e}\,e^{-\frac{\pi}{2 \lambda}}(1+\cal{O}(\lambda))$.

We now turn to the treatment of non-zero temperatures, following
refs \cite{millis96,ciuchi99}.  The crucial object in the analysis
is  the electron Greens function, which in
the $a-b$ basis introduced above is a  matrix ${\bf G}$ given by 
\begin{equation}
{\bf G}^{-1}_{ \vec{k},i\omega _{n}} =i\omega_n+\mu-\Sigma_{n}(i\omega_n)
-\epsilon_k {\bf \tau_x}-\Sigma_{co}(i\omega_n) {\bf \tau_z} \label{green}
\end{equation}
The self-energy $\Sigma$ has a normal (n) component at all $T$ and
and extra $co$ component at $T<T_{co}$. Its k-independence (locality)
is the fundamental assumption of the
dynamical mean field method.  Off diagonal $\sim {\bf \tau_x}$ components
of $\Sigma$ are not local in cordinate space, so do not appear in the DMFT.
The two self energies are fixed by
two coupled dynamical mean-field equations\cite{ciuchi99} 
for effective fields $(a,b)$:
\begin{eqnarray}
a_{n} &=&i\omega _{n}+\mu -\frac{W^{2}}{4}\,{\cal G_{BB}}\left( i\omega
_{n}\right) \;,  \label{dmft} \\
b_{n} &=&i\omega _{n}+\mu -\frac{W^{2}}{4}\,{\cal G_{AA}}\left( i\omega
_{n}\right) \;.
\end{eqnarray}
Here ${\cal G}_{AA/BB}$ are the two nonzero components of the 
local Greens function
defined by $G_{{\rm loc}}\left( i\omega \right) =1/N\sum_{\vec{k}}G\left( 
\vec{k},i\omega _{n}\right) $.(recall $\sum_k \epsilon_k=0$. 
They are related to the effective fields $\{a_{n},b_{n}\}$ by 
${\cal G}_{AA}\left(
i\omega _{n}\right) =\frac{{\rm \partial }\ln Z_{A}}{{\rm \partial }a_{n}}$
with $Z_{A}=\int\limits_{-\infty }^{\infty }dr\,P(\{a_{n}\},r)$ and 
\begin{equation}
P\left( \{a_{n}\},r\right) =e^{-\beta \left[ \frac{1}{2\lambda}
r^{2}-\left( r-\mu \right) n-\frac{1}{\beta }\sum_{n}\ln \left(
a_{n}-r\right) \right] }  \label{part}
\end{equation}
(${\cal G}_{B}$ and $Z_{B}$ analogous). $P\left( \{a_{n}\},r\right) $ is the
probability distribution of local distortions on the $A$-sublattice. 

For zero
temperature we recover the previously discussed commensurate charge order,
because for Eq. \ref{model}, at $T=0$ 
$\Sigma _{AA}\left( \vec{k}
,i\omega _{n}\right) =-\Delta $ and $\Sigma _{BB}\left( \vec{k},i\omega
_{n}\right) =\Delta $. For $T>0$ we solve Eq. (\ref{dmft}) to (\ref{part})
using the procedure of Ref. \onlinecite{millis96}; we retained $2^{10}$
Matusbara frequencies and extended the frequency range analytically. We
obtained convergence for  $T > 0.005$ which is roughly
$T_{co}$ for $\lambda =0.64$

\vspace{0.25cm}
\centerline{\psfig{file=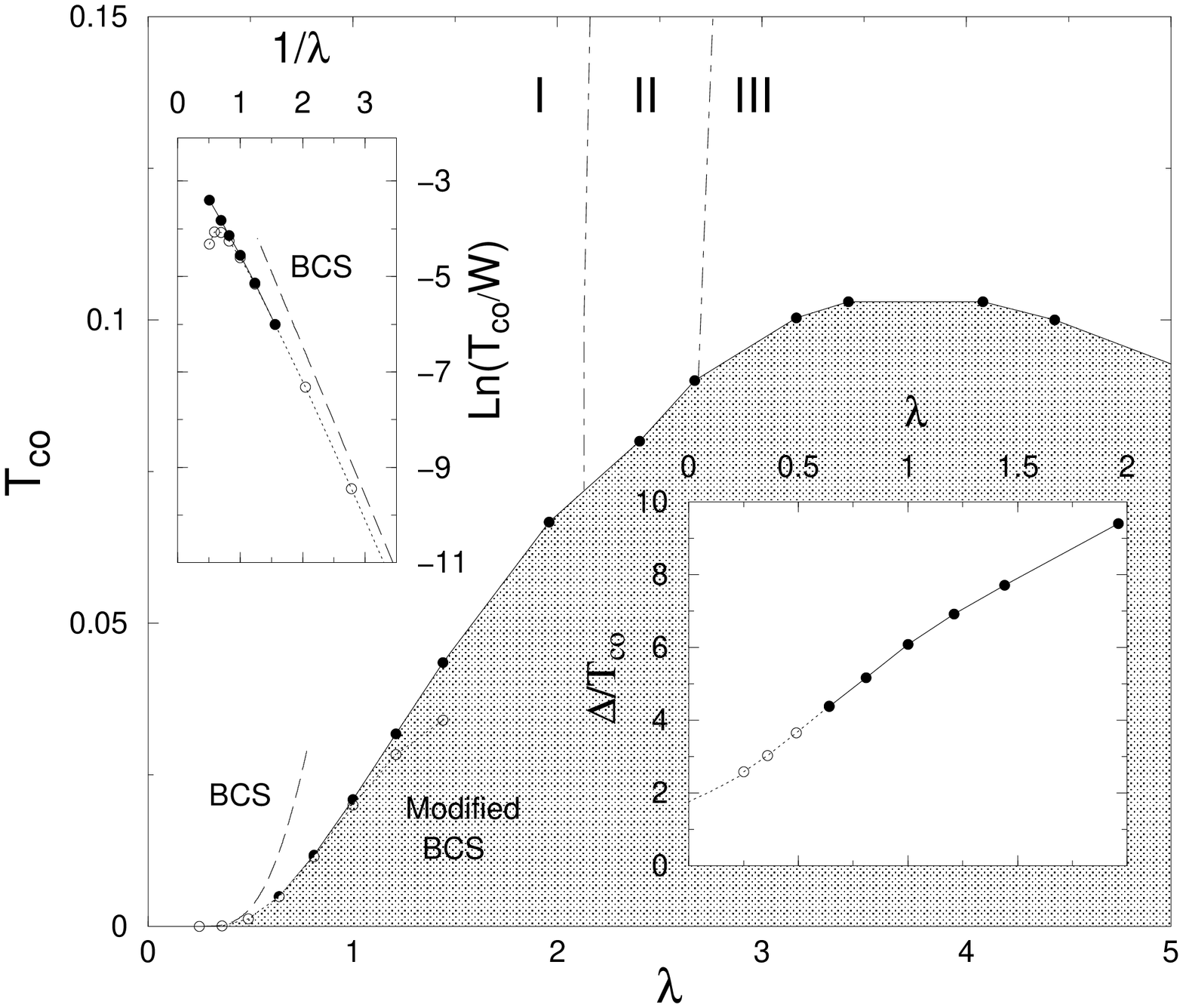,width=8cm,angle=270}}
{\footnotesize {\bf Fig 1} Main panel: phase diagram of 
$H_{hol}$. Solid points:  numerically
calculated transition temperatures. Open points:  'modified BCS'
approximation discussed below ({\ref{tco}}). Dashed line:  
conventional BCS approximation.
Shaded region: charge ordered; unshaded:  no order. Regime
I:  fermi liquid like normal state; regime III: 'polaron-like' normal
state; regime II cross-over, as defined from $P(r)$ and $\rho(\omega)$ 
as discussed in text below.
Left inset: $T_{co}$ vs $\lambda$ on logarithmic scale 
Right inset: gap to transition 
temperature ratio.}
\vspace{0.1in}

Fig. 1 shows our calculated phase diagram. 
Where there is overlap (roughly $\lambda >0.8$)
our results are in agreement with those of Ciuchi and de
Pasquale \cite{ciuchi99} if we scale $ T\rightarrow T/W$ and
$\lambda \rightarrow \lambda /(2W)$ with $W=2$.   From the
left inset we see that  
the BCS approximation (dashed line) overestimates
$T_c$ by a constant factor, but 
a different `modified BCS'  approximation 
(open circles) agrees well with the 
small $\lambda$ numerical results.  From the right
inset of Fig. 1 we see that
only for $\lambda = 0$ is the BCS value is obtained.

\vspace{0.30cm}

\centerline{\psfig{file=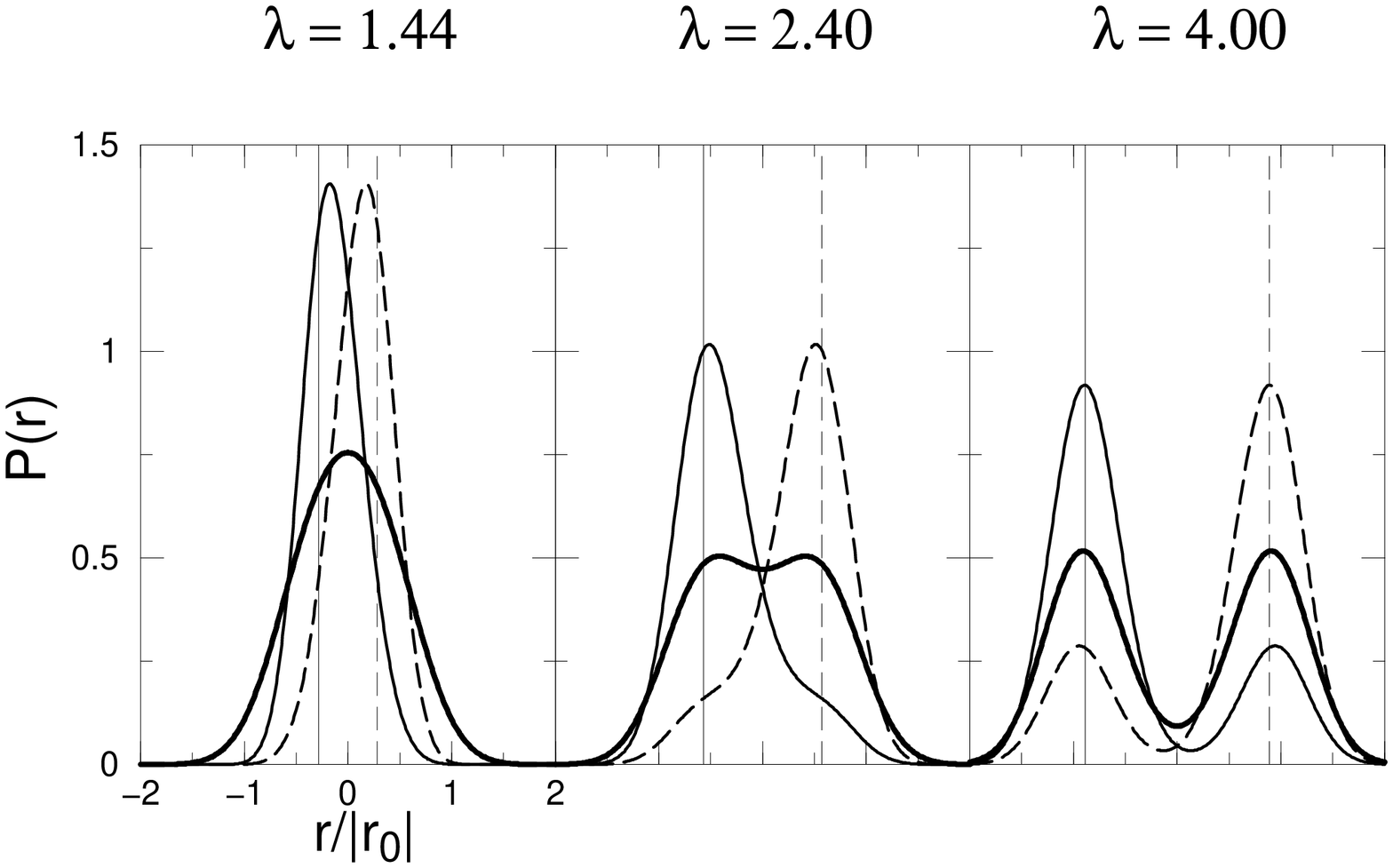,width=8cm,angle=270}}
{\footnotesize {\bf Fig. 2}
Distribution of local distortions $P\left(r\right)$
for three different coupling strengths $\protect\lambda$, for
$T=0$ (vertical line), $T \approx 0.93T_{co}$ (light lines) and
$T=0.125 >> T_{co}$ (heavy lines). For $T < T_{co}$
solid line:  A sublattice; dashed line:  B-sublattice.}
\label{fig:dis}

We next consider the phonon
probability distribution $P\left(r\right)$ and the local electron spectral
function $\rho\left(\omega\right) = -\frac{1}{\pi}\,{\rm Im}\,{\cal G}
_{A/B}\left(\omega\right)$, shown in Figs 2 and 3 for several 
couplings and temperatures.
 
\vspace{0.90cm}

\centerline{\psfig{file=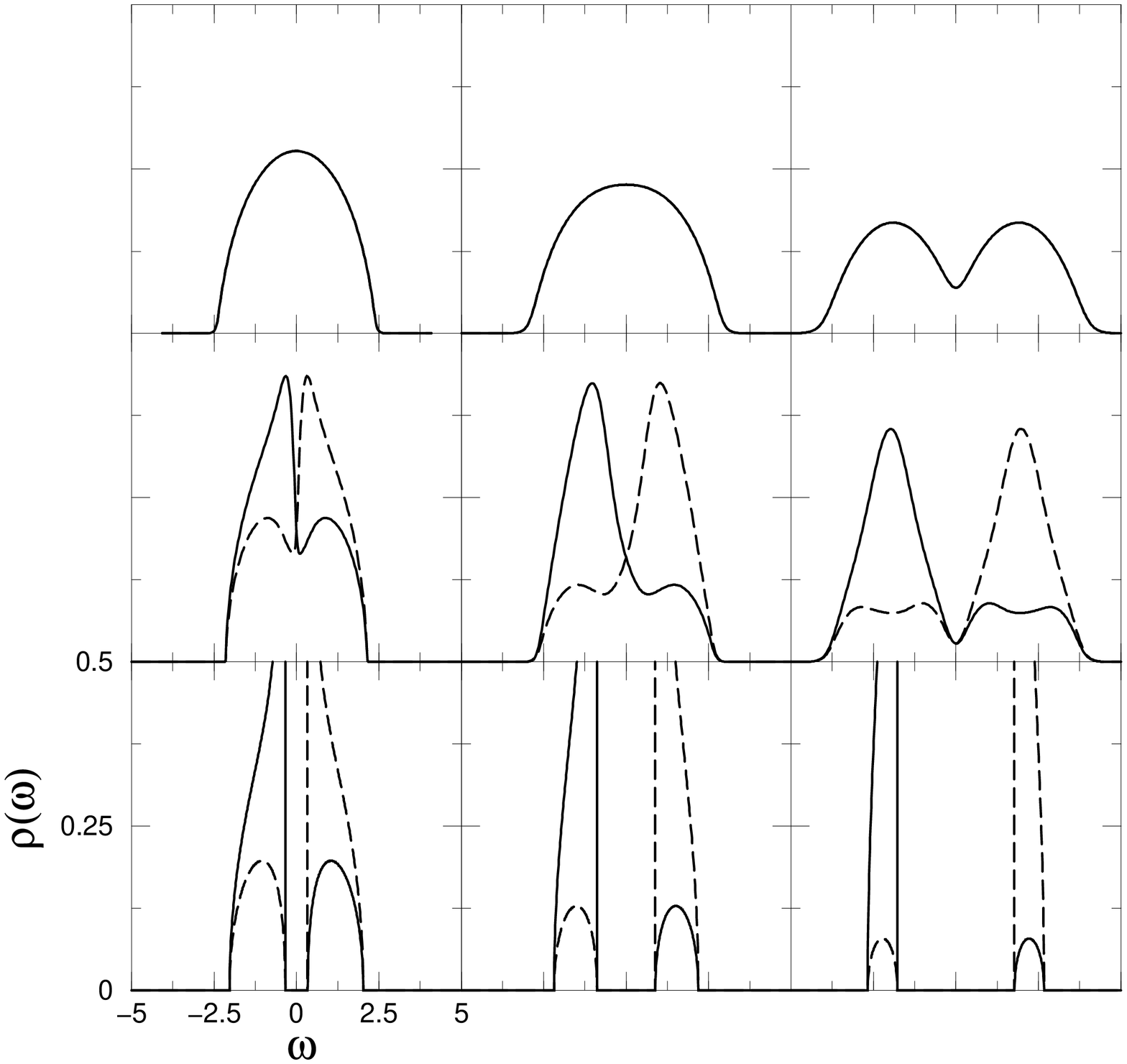,width=8cm,angle=270}}
{\footnotesize {\bf Fig. 3}  
Spectral function of electrons
$\protect\rho\left(\protect\omega\right)$ for three different coupling
strength $\protect\lambda$. for three different coupling strengths $\protect\lambda$, for
$T=0$ (vertical line), $T \approx 0.93T_{co}$ (light lines) and
$T=0.125 >> T_{co}$ (heavy lines). For $T < T_{co}$
solid line:  A sublattice; dashed line:  B-sublattice.}

\vspace{0.1in}

For a weakly coupled fermi liquid 
(left panels, Figs 2,3; region I of figure I) $P\left(r\right)$ 
is sharply peaked about its mean value $r=0$ and $
\rho\left(\omega\right)$ takes approximately
the non-interacting form (in our case, a
semicircle). 
For a very strongly coupled system (right panels, Figs 2,3;
region III, Fig. 1) one should think of the
excitations as polarons, strongly bound electron-phonon complexes. In this
situation $P\left(r\right)$ displays two broad peaks
(corresponding to distortions on occupied sites and antdistortions on
unoccupied sites) and $\rho\left(\omega\right)$ has a two peak
structure corresponding to bound and empty sites. 
The cross-over between regions
I and III occurs for
couplings near $\lambda_c = 3\pi/4 = 2.356$ at which the weak coupling
theory predicts a $T=0$ transition to a charge-disordered polaron insulator
\cite{millis96} (this transition is preempted by the charge ordering
transition). 
We have defined the left-hand boundary of the cross-over region as the locus
of points at which $P\left(r\right)$ loses its maximum at $r=0$ and the
right-hand boundary as the locus of points at which $\rho\left(\omega\right)$
gains a minimum at $\omega = 0$. The point 
$\lambda_c$ sits approximately in
the center of region II. 

We now examine  the weak coupling limit of $T_{co}$ analytically.  
We first assume the equations have
been solved for $T>T_{co}$, yielding a mean field function $c_{n}$. Just
below the transition temperature the two effective fields $a_{n}$ and $b_{n}$
differ only slightly from each other or\ from the $T>T_{co}$ result, thus $
a_{n}=c_{n}+\epsilon _{n}\,,\;\;\;b_{n}=c_{n}-\epsilon _{n}$. Expanding the
equations for the effective fields to linear order in $\epsilon _{n}$ we get
the matrix equation for $\epsilon_n$
\begin{equation}
\label{eps}
\epsilon _{n}=-\frac{W^{2}}{4}\,\left[ {\cal G}_{n}^{2}\,\epsilon
_{n}+\sum\limits_{m\neq n}\left( {\cal G}_{n}\,{\cal G}_{m}+\frac{{\cal G}
_{m}-{\cal G}_{n}}{c_{m}-c_{n}}\right) \,\epsilon _{m}\right] 
\end{equation}
with ${\cal G}_{n}=\frac{1}{Z}\,\int \,dr\,\frac{P\left(
\{c_{n}\},r\right) }{c_{n}-gr}$ .  $ T_{co}$ is the temperature at which 
Eq. \ref{eps} has a
solution. As written it is general;  
in the weak coupling limit ${\cal
G}_{n}=\frac{1}{c_{n}^{0}}\,\left( 1+\bar{\lambda}\,\frac{1}{\beta
}\,\frac{1}{(c_{n}^{0})^{2}}+{\cal O}\left(
T^{2}\bar{\lambda}^{2}\right) \right) $ with $c_{n}^{0}\;=\;\left(
{\cal G}_{n}^{0}\right) ^{-1}=\;\;\frac{1}{2}\,\left( i\omega
_{n}+\sqrt{\left( i\omega _{n}\right) ^{2}-W^{2}}\right) $ and
\begin{equation}
\bar{\lambda}=\lambda \,/\left( 1+\frac{\lambda W}{\beta }\,\sum\limits_{n}
\frac{1}{\left( c_{n}^{0}\right) ^{2}}\right) \;\;\equiv \;\;(\lambda^{-1}
-\lambda_c^{-1})^{-1}
\end{equation}
From the form of ${\cal G}_n$ we see that the expansion parameter is
$T\,\bar{\lambda}$ rather then $\lambda $. The factor $T$ is the
classical analogue of the Migdal parameter $\omega _{D}$ while the
presence of $\bar{\lambda}$ rather than $\lambda $ indicates that
proximity to the polaronic instability strongly renormalizes the
coupling. Inserting ${\cal G}_n$ into (\ref{eps})
and rearranging leads, after lengthy calculation to the weak-coupling
$T_{co}$-equation
\begin{equation}
1=-\lambda \,\frac{1}{\beta }\,\sum\limits_{n}\frac{{\cal
G}_{n}^{0}}{i\omega _{n}-2\,\bar{\lambda}\,T_{co}\,{\cal
G}_{n}^{0}}\;.  \label{tco}
\end{equation}

Eq.(\ref{tco}) may be evaluated to logarithmic
accuracy by replacing $\cal{G}$ by
its $\omega=0$ values $\pm 2/W$, yielding
$\frac{\Delta}{T_{\rm co}} = \frac{\pi}{\gamma} +
\frac{\pi^2}{\gamma}\,\frac{\bar{\lambda}}{W}$
($\gamma =1.78107$), showing explicitly the deviations from BCS.

 We now
compare Eq \ref{tco} to the result obtained from the 
divergence of the charge order
susceptibility defined by 
$\chi _{co} =\frac{1}{N}\,\sum\limits_{i}e^{-i\,\vec{Q}\vec{r}
_{i}}\,\int\limits_{0}^{\beta }\,d\tau \,\left( \langle \,n_{i}\left( \tau
\right) n_{0}\left( 0\right) \,\rangle - \langle \,n_{0}\left( 0\right) \,
\rangle \right)^2  \\
\equiv \left( \frac{1}{\beta }\right) ^{2}\sum\limits_{n,p}\chi _{np}$
$\chi_{np}$ may be expressed in terms of an irreducible
vertex function $\bar{\Gamma}$ which, 
within the dynamical mean-field approximation,
is local\cite{brandt89,zlatic90}. 
thus $\chi_{np}$ satisfies \cite{freericks94} 
$\chi _{np}=\chi _{n}^{0}\,\beta \,
\delta _{np}-\chi _{n}^{0}\,\frac{1}{\beta 
}\,\sum\limits_{m}\bar{\Gamma}_{nm}\,\chi _{mp}$
with $\chi _{n}^{0}=-\frac{1}{N}\sum\limits_{\vec{k}}G_{n}\left(
\vec{k}+\vec{Q}\right) G_{n}\left( \vec{k}\right) \;=\;-\frac{{\cal
G}_{n}}{i\omega _{n}-\Sigma _{n}}$. This equation may be expressed
diagrammatically; in
the weak coupling limit, the terms are  shown in Fig. 4, which
is  exact up to terms of order $\left(T\lambda \right) ^{2}$. 

\vspace{0.25cm}
\centerline{\psfig{file=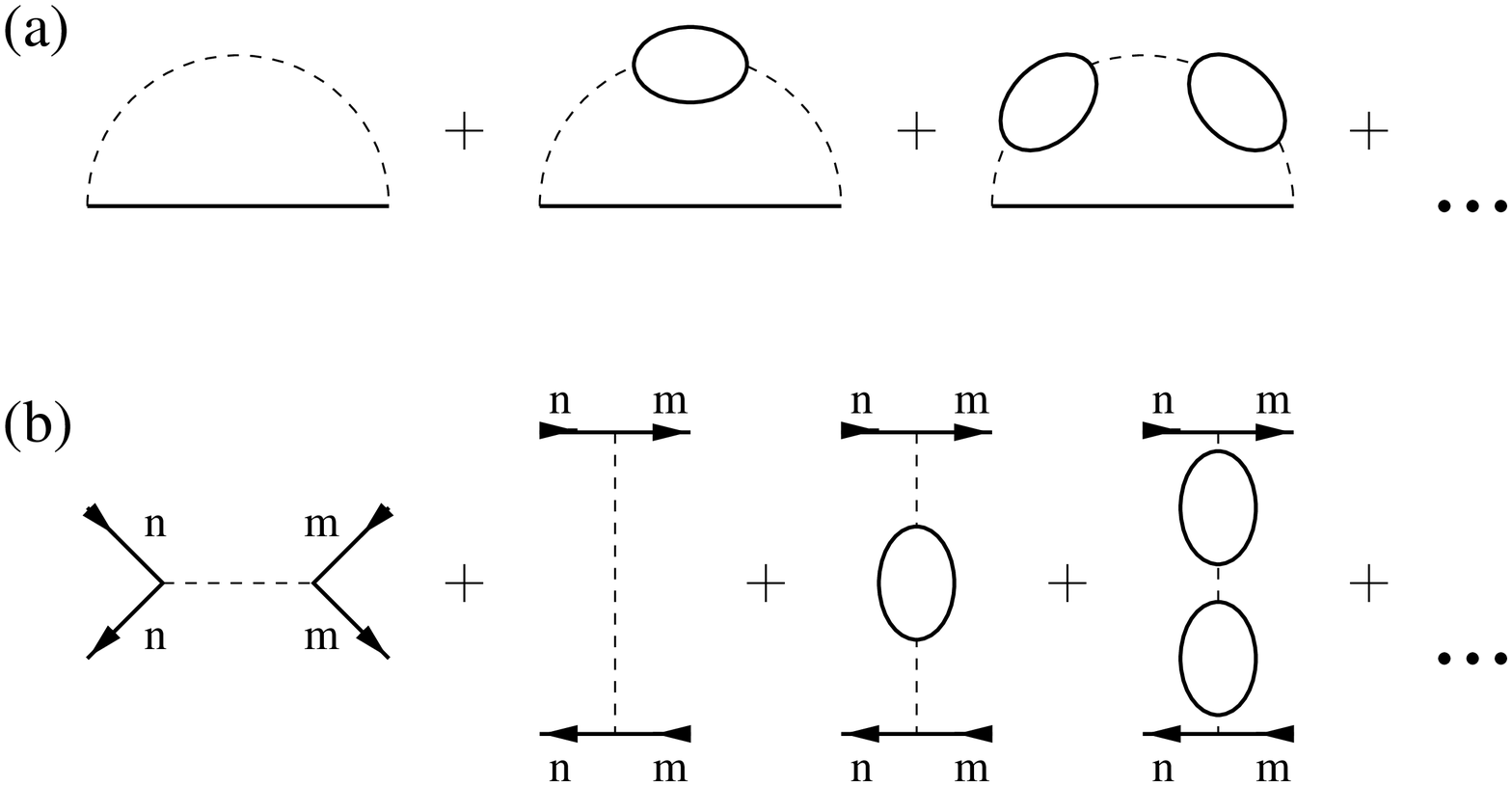,width=8cm,angle=0}}
\vfill
{\footnotesize {\bf Fig. 4 }
Diagrammatic expansion of the one-particle self-energy (a) and of
the irreducible vertex of the charge-order susceptibility (b). 
Heavy lines: full local Green's function ${\cal G}_n$. Dashed
lines: free phonon propagator $D^0_n$. All not shown but
possible diagrams are smaller by additional factors of $T$.}

\vspace{0.1in}

We find $\Sigma _{n} =\bar{\lambda}\,T $ and 
$\bar{\Gamma}_{nm} =-\lambda +\bar{\lambda}\,\delta _{nm}$.
with $\bar{\lambda}=\lambda \,/( 1+\lambda \,
\frac{1}{\beta }\,\sum
{\cal G}_{n}^{2})$.
Note, $\bar{\lambda}$ coincides with the previously introduced one if we
replace ${\cal G}_{n}\rightarrow {\cal G}_{n}^{0}$, which is
correct at the order in $T$ to which we work.  Use of
this $\bar{\Gamma}$ yields  $\chi ^{-1}=\tilde{\chi}
_{0}^{-1}-\lambda $ with $\tilde{\chi}_{0}=\frac{1}{\beta }\,\sum\limits_{n}
\frac{\chi _{n}^{0}}{1+\bar{\lambda}\,T\,\chi _{n}^{0}},$. This
diverges when Eq. \ref{tco} is fulfilled.

We can now interpret Eq.(\ref{tco}) on more physical grounds. In the
BCS limit, the irreducible vertex is given by $\lambda $ and $\chi
_{n}^{0}$ by the convolution of two bare Green's function. 
The leading effect of increased coupling
is a $T$-dependent self-energy, corresponding to inelastic scattering
of electrons by phonons.  Insertion of self-energies into $\chi
_{n}^{0}$ increases the scale at which the logarithmic divergence is
cut off, thereby reducing $T_{co}$. This physics however does not
operate at $T=0$; thus it does not reduce the gap.  In a more
realistic model including quantum phonons some inelastic scattering
would be present at $\omega >0$ even at $T=0$, renormalizing the gap
slightly, but it would be weak in comparison to the thermally generated
scattering so $T_{co}$ will be reduced more then $\Delta $. Also in a
more realistic quantum model the thermal scattering is negligible for
temperatures lower than $\Theta _{{\rm D}}/3$ ($\Theta _{{\rm D}}$ is
the Debye temperature); thus the effects we have described
will be important only if $T_{co} {> \atop \sim } \Theta _{{\rm
D}}/3$. In
addition to the self-energy, there is a vertex correction. which acts
just as the self-energy, to weaken the instability via inelastic
scattering.

The 'modified BCS' result \ref{tco} is shown as the dashed line in
Fig. 1 and is seen to break down at the boundry of region I, when
$\rho\left(\omega\right)$ and $P\left(r\right)$ begin to
acquire structure
indicating polaronic features. This shows that in the weak coupling
regime, inelastic scattering is the key issue, and that 'pseudogap'
formation not easily described in terms of traditional diagrammatics
leads to changes in behavior.

We finally discuss the relation of our results to the
interesting proposal of  McMillan \cite{mcmillan77}
concerning phonon entropy effects. 
As can be seen from Fig. 2, 
at intermediate coupling thermal fluctuations of
the phonon fields are crucial. If the fluctuations were negligible as
in the BCS theory the widths of the peaks at $T=0.04$ would be
negligible in comparison to their separation. However, as can also be
seen the peaks either in the phonon distribution or the density of
states do not shift appreciably until $T$ becomes extremely close to
$T_{co}$. Thus the fluctuations, while strong,
apparently do not feed back and reduce the mean
value of the gap in the manner assumed by McMillan. However,
the fluctuations do lead to a non-negligible probability of $r \approx
0$, i.e., of regions with a very small local gap. Electronic
excitations in these regions give rise to the non-vanishing low energy
density of states at the chemical potential, and as the regions become
more probable, cause the destruction of the ordered phase. Comparison
of the midlle and left panel of Figs.2 and 3  
shows that as the coupling is further increased, the
physics crosses over smoothly to that expected in the very strong
coupling limit.

{\it Acknowledgements} We thank 
A.~Rosch and G.~Chitov for useful discussions. SB 
acknowledges 
the DFG and the Rutgers University Center for Materials Theory for
financial support; AJM acknowledges NSF DMR9906995.


\end{multicols}
\end{document}